\documentclass[a4paper,10pt,twoside]{cpc-hepnp}

\usepackage{multicol}
\usepackage{graphicx}
\usepackage{booktabs}
\usepackage{amssymb,bm,mathrsfs,amscd}
\usepackage[tbtags]{amsmath}

\begin{document}

\fancyhead[co]{\footnotesize H. Kamano: Study of excited nucleon states at EBAC: status and plans}

\footnotetext[0]{Received 20 June 2009}

\title{Study of excited nucleon states at EBAC:\\ status and plans\thanks{
This work is supported by 
the U.S. Department of Energy, Office of Nuclear Physics Division, under 
contract No. DE-AC02-06CH11357, and Contract No. DE-AC05-06OR23177 
under which Jefferson Science Associates operates Jefferson Lab.
Notice: Authored by Jefferson Science Associates, LLC under U.S. 
DOE Contract No. DE-AC05-06OR23177. 
The U.S. Government retains a non-exclusive, paid-up, irrevocable, world-wide 
license to publish or reproduce this manuscript for U.S. Government purposes. 
}}

\author{Hiroyuki Kamano$^{1;1)}$\email{hkamano@jlab.org}}

\maketitle

\address{
1~(Excited Baryon Analysis Center (EBAC), Thomas Jefferson National Accelerator Facility, Newport News, VA 23606, USA)
}

\begin{abstract}
We present an overview of a research program for 
the excited nucleon states in Excited Baryon Analysis Center (EBAC) 
at Jefferson Lab.
Current status of our analysis of the meson production reactions
based on the unitary dynamical coupled-channels model is summarized, 
and the $N^\ast$ pole positions extracted from the constructed 
scattering amplitudes are presented.
Our plans for future developments are also discussed.
\end{abstract}



%
%
\begin{multicols}{2}
\section{\label{sec:intro}Introduction}
Understanding the structure of hadrons and their excitations still remains
as an outstanding and fundamental challenge in the hadron physics.
The existence of the hadrons is a direct consequence of the confinement
of quarks and gluons within the quantum chromodynamics (QCD).
Thus exploring the hadron structure is
nothing but making clear the non-perturbative nature of QCD.

Recent experimental activities at electron-beam facilities 
such as JLab, Bonn, Mainz, SPring-8, and GRAAL have brought
new light to the study of excited nucleon ($N^\ast$) states\cite{Bur04}.
Those facilities provide a considerable amount of precise data of the
meson photo- and electro-production reactions on the nucleon target,
and open a great opportunity to make quantitative investigations of the
$N^\ast$ structure.

Against the background of those experimental progress, continuous effort
to extract the $N^\ast$ properties from the world data of meson
production $\pi N$, $\gamma N$, and $N(e,e')$ reactions
is being made in Excited Baryon Analysis Center (EBAC) at Jefferson Lab.
The analysis is pursued with a dynamical coupled-channels (DCC)
approach proposed in Ref.~\citep{msl07}.
The approach treats most relevant unitary cuts below $W=2$~GeV,
i.e., $\pi N$, $\eta N$, and $\pi\pi N$ which has 
unstable $\pi\Delta$, $\rho N$, $\sigma N$ channels, 
in solving the coupled-channels equations.

The objective of EBAC is more than just performing
the partial-wave analysis of the meson production reactions.
We not only try to extract the $N^\ast$ parameters,
but also to map out the quark-gluon substructure of the $N^\ast$ states.
It will require comprehensive study combined with
various hadron structure calculations such as constituent quark models,
covariant models based on Dyson-Schwinger equations, 
and Lattice QCD simulations.
The $N^\ast$ information extracted from the reaction model analysis,
such as that performed in EBAC,
is vital for bridging the gap between 
the actual reaction data and the hadron structure calculations.

The main subjects in EBAC are summarized as follows:
\begin{enumerate}
\item Establish baryon spectrum.
\item Extract $N^\ast$ parameters, in particular the electromagnetic
$N$-$N^\ast$ transition form factors,
from analyzing the world data of meson production 
$\pi N$, $\gamma N$, $N(e,e')$ reactions.
\item Develop a method to connect the extracted form factors to 
the hadron structure calculations and deduce the structure of 
the $N^\ast$ states.
\end{enumerate}

In this paper we present an overview of current status and 
future plans of the EBAC program.
In Sec.~{\ref{sec:ebac-dcc}} we briefly explain a framework of
the EBAC-DCC model,
and present the results of our analysis
of the meson production reactions in Sec.~{\ref{sec:analysis}}.
In Sec.~{\ref{sec:pole}} we present the extracted $N^\ast$ pole positions
within the current EBAC-DCC model.
Finally we discuss our future plan in Sec.~{\ref{sec:plan}}.
%
%
%
\section{\label{sec:ebac-dcc}EBAC-DCC model}
The EBAC-DCC analysis is based on a
multi-channels and multi-resonances model\cite{msl07}
within which the partial wave amplitudes of
$M(\vec p) + B(-\vec p) \to M'(\vec p') + B'(-\vec p')$
are calculated from the following coupled-channels integral 
equation (suppressing the angular momentum and isospin indices):
\begin{eqnarray}
T_{MB,M'B'}(p,p';E) &=& 
V_{MB,M'B'}(p,p';E)
+
\nonumber\\
&&
\!\!\!\!\!\!\!\!\!\!\!\!\!\!\!
\!\!\!\!\!\!\!\!\!\!\!\!\!\!\!
\!\!\!\!\!\!\!\!\!\!\!\!\!\!\!
\sum_{M''B''}\int^\infty_0 dq q^2 V_{MB,M''B''}(p,q;E)
\times
\nonumber\\
&&
\!\!\!\!\!\!\!\!\!\!\!\!\!\!\!
\!\!\!\!\!\!\!\!\!\!\!\!\!\!\!
\!\!\!\!\!\!\!\!\!\!\!\!\!\!\!
G_{M''B''}(q;E) T_{M''B'',M'B'}(q,p';E).
\label{eq:lseq}
\end{eqnarray}
Here $V_{MB,M'B'}$ and $G_{MB}$ are the
$MB\to M'B'$ transition potential and the $MB$ Green function
described below; the $M''B''$ summation represents 
the coupled-channels effects among the
$\pi N,\eta N,\pi\Delta,\rho N,\sigma N$ channels
in the reaction processes.
It is well known that the solution of the above equation
automatically satisfies the unitary conditions for the scattering
amplitudes.

The $MB\to M'B'$ transition potential is defined as
\begin{eqnarray}
V_{MB,M'B'}(p,p';E) 
&=& 
v_{MB,M'B'}(p,p') 
+ 
\nonumber\\
&&
\!\!\!\!\!\!\!\!\!\!\!\!\!\!\!
\!\!\!\!\!\!\!\!\!\!\!\!\!\!\!
\sum_{N^\ast_i}
\dfrac{\Gamma^\dag_{N^\ast_i,MB}(p)\Gamma_{N^\ast_i,M'B'}(p')}{E-m^0_{N^\ast_i}},
\label{eq:pot}
\end{eqnarray}
where $m^0_{N^\ast_i}$ and $\Gamma^\dag_{N^\ast_i,MB}(p)$
represent the bare mass of the $i$th $N^\ast$ state and
the bare $N^\ast_i \to MB$ decay vertex, respectively.
The meson exchange potential $v_{MB,M'B'}$ is derived using
the unitary transformation method\cite{sl96,sl09} from the phenomenological
Lagrangians which respects gauge and chiral symmetries.
It is noted that the derived potential is independent of
the total scattering energy $E$.
The second term describes a $MB\to M'B'$ transition through the propagation of 
the bare $N^\ast$ state, $MB\to N^\ast\to M'B'$.
Also, defining  $E_\alpha(k)=[m^2_\alpha + k^2]^{1/2}$ with $m_\alpha$ being
the mass of particle $\alpha$, 
the meson-baryon propagators in Eq.~(\ref{eq:lseq}) are
$G_{MB}(k,E)=1/[E-E_M(k)-E_B(k) + i\epsilon]$ for the stable 
$\pi N$ and $\eta N$ channels,
and $G_{MB}(k,E)=1/[E-E_M(k)-E_B(k) -\Sigma_{MB}(k,E)]$
for the unstable $\pi\Delta$, $\rho N$, and $\sigma N$ channels. 
The self energy
$\Sigma_{MB}(k,E)$ is calculated from a vertex function defining the decay of
the considered unstable particle in the presence of a spectator $\pi$ or $N$ 
with momentum $k$. For example, we have for the $\pi\Delta$ state,
\begin{eqnarray}
\Sigma_{\pi\Delta}(p,E) &=&\frac{m_\Delta}{E_\Delta(p)}
\int^\infty_0 q^2 dq \frac{ M_{\pi N}(q)}{[M^2_{\pi N}(q) + p^2]^{1/2}}
\times
\nonumber \\
&&
\!\!\!\!\!\!\!\!\!
\frac{\left|f_{\Delta \to \pi N}(q)\right|^2}
{E-E_\pi(p) -[M^2_{\pi N}(q) + p^2]^{1/2} + i\epsilon},
\label{eq:self-pid}
\end{eqnarray}
where $M_{\pi N}(q) =E_\pi(q)+E_N(q)$ and $f_{\Delta \to \pi N}(q)$
defines the decay of the $\Delta \to \pi N$ in the rest frame of $\Delta$. 
The self-energies for $\rho N$ and $\sigma N$ channels are similar.

We can split the full partial wave amplitude $T_{MB,M'B'}(p,p';E)$
into two pieces without introducing any approximation:
\begin{eqnarray}
T_{MB,M'B'} (p,p';E) &=& t_{MB,M'B'} (p,p';E) + 
\nonumber\\
&&
t^{N^\ast}_{MB,M'B'} (p,p';E).
\label{eq:decomp}
\end{eqnarray}
Here the first term is a solution of Eq.~(\ref{eq:lseq})
but replacing the full transition potential $V_{MB,M'B'}$ with
the meson-exchange potential $v_{MB,M'B'}$.
Therefore $t_{MB,M'B'} (p,p';E)$ expresses
the pure meson-exchange processes
and thus is called the meson-exchange amplitude in this paper.
The second term describes reaction processes associated with
the bare $N^\ast$ states,
\begin{eqnarray}
t^{N^*}_{MB,M^\prime B^\prime}(p,p';E)&=& \sum_{N^*_i, N^*_j}
\bar{\Gamma}_{MB,N^*_i}(p;E) 
\times
\nonumber \\
&&
\!\!\!\!\!\!\!\!\!\!\!\!\!\!\!
[D(E)]_{i,j}
\bar{\Gamma}_{N^*_j, M^\prime B^\prime}(p';E),
\label{eq:tmbmb-r}
\end{eqnarray}
where $\bar{\Gamma}_{N^*_j, M' B'}(k;E)$ is the
dressed $N^\ast\to M'B'$ vertex function which is defined as
\begin{eqnarray}
\bar{\Gamma}_{N^*_j,M'B'}(p,E)
&=&
\Gamma_{N^*_j,M'B'}(p)+
\nonumber\\
&&
\!\!\!\!\!\!\!\!\!\!\!\!\!\!\!
\!\!\!\!\!\!\!\!\!\!\!\!\!\!\!
\!\!\!\!\!\!\!\!\!\!\!\!\!\!\!
\sum_{M''B''}
\int dq q^2
\Gamma_{N^*_j, M''B''}(q)
G_{M''B''}(q;E)
\times
\nonumber\\
&&
t_{M''B'',M'B'}(q,p;E).
\end{eqnarray}
The inverse of the propagator of dressed $N^*$ states in
Eq.~(\ref{eq:tmbmb-r}) is 
\begin{eqnarray}
[D^{-1}(E)]_{i,j} &=& (E - m^0_{N^*_i})\delta_{i,j} - [M(E)]_{i,j},
\label{eq:nstar-selfe}
\end{eqnarray}
where the $N^*$ self-energy is defined by
\begin{eqnarray}
[M(E)]_{i,j}&=&
\sum_{MB}
\int^\infty_0 \!\!\!\! q^2 dq 
\bar{\Gamma}_{N^*_j \to M B}(q,E)
\times
\nonumber \\ 
&&
G_{MB}(q,E)\,
{\Gamma}_{MB \to N^*_i}(q,E).
\label{eq:nstar-g}
\end{eqnarray}

As for the inclusion of the bare $N^\ast$ states, we impose a condition
that the number of the bare $N^\ast$ states included should be 
as minimum as possible.
This means that if we can explain the data even without 
some of the bare $N^\ast$ states, then we will take those states out of our
framework.
On the other hand, if we cannot explain some data set with 
the current framework,
then we will try to introduce an additional bare $N^\ast$ state and 
it may correspond to the new $N^\ast$ state.
At present we have introduced 16 number of the bare $N^\ast$ states.

As clearly seen in Eqs.~(\ref{eq:tmbmb-r})-(\ref{eq:nstar-g}),
the bare $N^\ast$ states couple to the meson-baryon continuum
through the reaction processes and become resonance states.
On the other hand, the meson-exchange amplitude, which describes the pure 
meson-exchange processes, can also generate
the resonance poles dynamically.
Our model allows both possibilities.

The $MB\to M'B'$ amplitude defined in Eqs.~(\ref{eq:lseq})-(\ref{eq:nstar-g})
constitutes a basic ingredient to construct
all single and double meson production reactions
with the initial $\pi N$ and $\gamma^{(\ast)} N$ channels.
The details of constructing the $\pi N\to \pi\pi N$ and 
$\gamma^{(\ast)} N\to MB,\pi\pi N$ amplitudes
can be found in Refs.~\citep{msl07,kjlms09}.
%
%
%
\section{\label{sec:analysis}Current status of the dynamical coupled-channels analysis at EBAC}
\subsection{$\pi N\to\pi N$ scattering}
As steps toward extracting information on the $N^\ast$ states,
it is necessary to first determine the hadronic parameters 
of the EBAC-DCC model.
This was accomplished in Ref.~\citep{jlms07}.
There the SAID $\pi N$ partial wave amplitudes 
are considered as the ``experimental data''
and the hadronic parameters are determined by fitting to them up to $W=2$ GeV.
The numerical fit was performed systematically using the MINUIT library.
We refer to this analysis and the obtained parameter set as JLMS
in the following.

In Fig.~\ref{fig:pin-sig-p} our results of the 
angular distributions and the polarizations 
of $\pi^-p \to\pi^-p$ and $\pi^-p\to\pi^0n$ calculated
from the EBAC-DCC model with the JLMS parameters (red solid curves) are
compared with the SAID results (blue dashed curves).
Our results agree with the SAID results as well as the data.
As a next step we plan to improve our hadronic parameters 
by fitting to the data of these observables.

In Fig.~\ref{fig:1pi-tot}, we also present the resulting total cross 
sections of $\pi N\to X$ (red solid curves) and 
$\pi N\to \pi N$ (blue dashed curves) which agree with the data well.
The deviation of the $\pi N\to X$ total cross sections above $W=1.8$ GeV
can be understood because 
in those energy region the multi-pion production reactions,
which are not considered in the current EBAC-DCC model, 
also start to visibly contribute to the total cross sections. 
\end{multicols}

\ruleup
\begin{center}
\includegraphics[clip,height=6cm]{fig1-1.eps}~~~~
\includegraphics[clip,height=6cm]{fig1-2.eps}
\figcaption{
\label{fig:pin-sig-p}
Differential cross sections (left) and polarizations (right)
of $\pi^- p \to \pi^- p$ and $\pi^- p\to \pi^0 n$.
The results from EBAC-DCC analysis~\citep{jlms07} (red solid curves) 
are compared with those from GWU group~\citep{said06} (blue dashed curves).
}
\end{center}
\begin{center}
\includegraphics[clip,width=13cm]{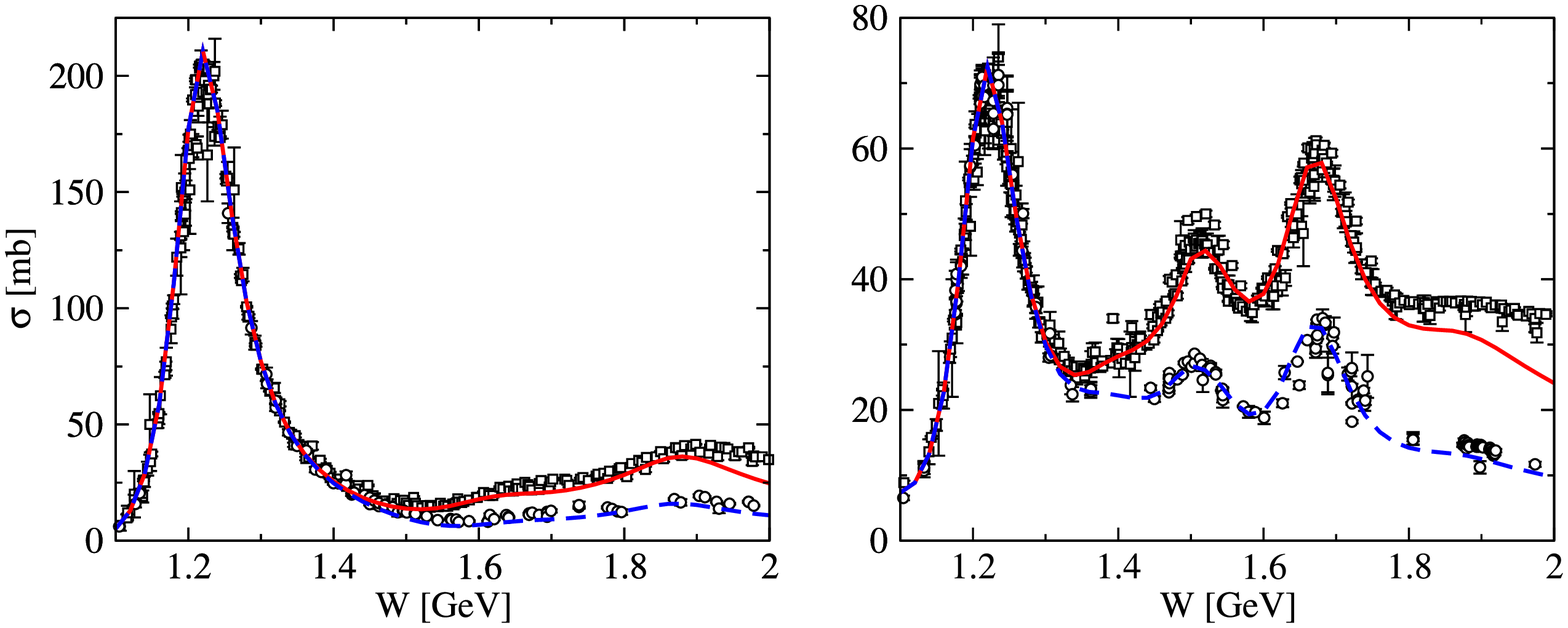}
\figcaption{
\label{fig:1pi-tot}
The predicted $\pi N$ total cross sections from the EBAC-DCC model.
(Left panel) 
The $\pi^+ p \to X$ (solid red curve) 
and $\pi^+ p \to \pi^+ p$ (dashed blue curve) reactions.
(Right panel) 
The $\pi^- p \to X$ (solid red curve) and 
$\pi^- p \to \pi^- p+\pi^+ n$ (dashed blue curve) reactions.
The corresponding data to to the solid and dashed curves are 
presented as open square and open circle, respectively.
The data are from Refs.~\citep{pdg,gwuweb}
}
\end{center}
\ruledown

\begin{multicols}{2}
\subsection{$\pi N\to\pi\pi N$ reaction}
With the JLMS parameters obtained from the analysis of the $\pi N$ scattering, 
we have calculated cross sections of the $\pi N\to \pi\pi N$ reactions
in Ref.~\citep{kjlms09}.
Within our model there appears no additional parameters in constructing
the $\pi N\to\pi\pi N$ amplitude. 
Thus the results are purely our predictions.

Figure.~\ref{fig:p2ptcs} is the resulting total cross sections 
(red solid curves).
Without any modifications of the parameters, our results already capture 
the essential features of the total cross sections up to $W=2$ GeV.
One possible reason for the deviation between our results and the data 
will be because the $\pi N$ elastic scattering does not completely fix 
the parameters associated with the inelastic 
$\pi\Delta$, $\rho N$, and $\sigma N$ channels.
This result indicates that we need a combined, simultaneous
analysis of the $\pi N$ elastic and $\pi N\to \pi\pi N$ reactions
to construct more complete hadronic amplitudes up to $W=2$ GeV.
In the same figure, we also present the results in which the coupled-channels
effect is turned off.
We observe that the coupled-channels effect is significant in all $W$ region.
\end{multicols}

\ruleup
\begin{center}
\includegraphics[clip,width=13cm]{fig3.eps}
\figcaption{\label{fig:p2ptcs} The predicted $\pi N\to \pi\pi N$
total cross sections. Red solid curves are the full results,
and blue dashed curves are the results in which the coupled-channels
effect is turned off.
See Ref.~\citep{kjlms09} for the data references.
}
\end{center}
\ruledown

\begin{multicols}{2}
In Fig.~\ref{fig:p2pinvms}, we present the invariant mass distributions
of $\pi^- p\to \pi^+\pi^- n$.
The shape of our results agrees with the data very well except
for the $\pi^+\pi^-$ distribution at low $W$.
It is expected that this disagreement with 
the data provides useful information to
refine the parameters associated with the coupling of the low lying $N^\ast$
states to the $\rho N$ and $\sigma N$ channels.
\begin{center}
\includegraphics[clip,width=8cm]{fig4.eps}
\figcaption{
\label{fig:p2pinvms}
The predicted invariant mass distributions
of the $\pi^- p\to \pi^+\pi^- n$ reaction 
at $W=1.44,~1.66$, and $1.79$ GeV.
The red solid curves are the full results, the dotted curves
are the phase-space normalized to the data.
The data are from Ref.~\citep{arndt}, whose magnitude is determined by
normalizing them to the $\pi^- p\to\pi^+\pi^- n$ total cross sections
listed in Ref.~\citep{manley}.
}
\end{center}
\subsection{$\gamma N\to\pi N$ and $e N\to e'\pi N$ reactions}
The main purpose for analyzing single pion photo- and electro-production
reactions is to extract the $N$-$N^\ast$ electromagnetic transition 
form factors.
Their precise determination is crucial for 
understanding the $N^\ast$ structure because
their $Q^2$ dependence is expected to strongly reflect
nature of the $N^\ast$ structure.

Within the EBAC-DCC model the $N$-$N^\ast$ electromagnetic transition 
form factors are obtained from the dressed $\gamma^\ast N\to N^\ast$
vertex function defined by
\begin{eqnarray}
\bar \Gamma_{\gamma^{(\ast)} N,N^\ast} (q,Q^2;E) &=&
\Gamma_{\gamma^{(\ast)} N,N^\ast} (q,Q^2)+
\nonumber\\
&&
\!\!\!\!\!\!\!\!\!\!\!\!\!\!\!
\!\!\!\!\!\!\!\!\!\!\!\!\!\!\!
\sum_{M''B''}\int dk k^2
v_{\gamma^{(\ast)} N,M''B''}(q,k,Q^2) 
\times
\nonumber\\
&&
\!\!\!\!\!\!\!\!\!\!\!\!\!\!\!
\!\!\!\!\!\!\!\!\!\!\!\!\!\!\!
G_{M''B''}(k;E)
\bar \Gamma_{M''B'',N^\ast} (k;E),
\end{eqnarray}
where $\Gamma_{\gamma^{(\ast)} N,N^\ast} (q,Q^2)$ is a bare 
$\gamma^{(\ast)} N\to N^\ast$ vertex function and
$v_{\gamma^{(\ast)}N,M''B''}(k,q,Q^2)$ is a meson-exchange 
$\gamma^{(\ast)}N\to M''B''$ transition potential.

Our first analysis of the $\gamma p\to\pi N$
and $p(e,e'\pi)N$ reactions has been performed up to $W\leq 1.6$ GeV 
in Refs.~\citep{jlmss08} and~\citep{jklmss09}, respectively.
In the analysis we fixed the hadronic parameters with the JLMS values 
and varied only the parameters associated with electromagnetic interactions,
i.e., $\Gamma_{\gamma^{(\ast)} N,N^\ast} (q,Q^2)$ and 
$v_{\gamma^{(\ast)} N,MB}(k,q,Q^2)$.

Figures~\ref{fig:g1ptcs} and~\ref{fig:5dim}
present our results of the 
total cross sections of $\gamma p\to \pi N$ and
the five-fold differential cross sections of $p(e,e'\pi)N$ 
at $Q^2=0.4$ (GeV/c)$^2$, respectively (red solid curve).
Our results agree with the data very well in the considered energy region
up to $W=1.6$ GeV.
We also present that the results in which 
the coupled-channels effect on the electromagnetic interactions 
is turned off (blue dashed curves).
As for the photoproduction reactions,
we find that the coupled-channels effect has about 30-40\% 
of contributions to the cross sections in all $W$ region up to 1.6 GeV.
On the other hand, the coupled-channels effect on 
the electroproduction reactions is still large around $W=1.2$ GeV  
but becomes small rapidly at high $W$ for increasing $Q^2$.

The resulting $G_M^\ast$, $G_E^\ast$, and $G_C^\ast$ form factors
of $\gamma^\ast N\to \Delta(1232)$ and the helicity amplitudes
for the higher $N^\ast$ states evaluated at their Breit-Wigner masses
are found in Refs.~\citep{jlmss08} and~\citep{jklmss09}.
\end{multicols}

\ruleup
\begin{center}
\includegraphics[clip,width=13cm]{fig5.eps}
\figcaption{\label{fig:g1ptcs} 
The total cross sections of 
$\gamma p\to\pi^0 p$ (left) and $\gamma p\to\pi^+ n$ (right).
The red solid curves are the full results of EBAC-DCC model;
the blue dashed curves are the results in which the coupled-channels
effect in the electromagnetic interactions is turned off 
(see Ref.~\citep{jlmss08} for the details).
}
\end{center}
\begin{center}
\includegraphics[clip,width=12cm]{fig6.eps}
\figcaption{\label{fig:5dim} 
The five-fold differential cross sections
of $p(e,e'\pi^0)p$ (upper panels) 
and $p(e,e'\pi^+)n$ (lower panels) at $Q^2 = 0.4$ (GeV/c)$^2$.
The meaning of each curve is same as in Fig.~\ref{fig:g1ptcs}.
The results are taken from Ref.~\citep{jklmss09}.
}
\end{center}
\ruledown

%
%
\begin{multicols}{2}
\section{\label{sec:pole}Extraction of $N^\ast$ poles from 
the constructed scattering amplitudes}
Once the scattering amplitude is constructed from analyzing 
the data of meson production reactions, we can extract various 
information on the $N^\ast$ states such as masses, widths,
and decay vertex functions.
The general scattering theory tells us that the $N^\ast$
mass and width should be identified with a pole position
of the scattering amplitude on the complex energy plane
and the $N^\ast$ decay vertex functions
with the residues at the corresponding pole\cite{goldberger,madrid02,bohm05}.

However, the scattering amplitude, which is
obtained as a solution of the coupled-channels equation~(\ref{eq:lseq}),
is originally not defined on the 
complex energy region where the resonance poles exist.
To explore the $N^\ast$ pole positions, we need to make an analytic
continuation of the scattering amplitudes.
How to perform such analytic continuation
both mathematically and numerically is 
described in detail in Ref.~\citep{ssl09}.
Therefore here we just present the results of the $N^\ast$ pole positions 
extracted from our constructed scattering amplitude.

In Table~\ref{tab1}, the extracted $N^\ast$ pole positions
from our current EBAC-DCC model are compared with the PDG values. 
It is found that most of our results and the PDG values are close to
each other.
It is noted, however, that we obtain these pole positions from just analyzing
the $\pi N$ scattering and did not try to make our results closed to the PDG
values at all.

Within our current model, all the extracted $N^\ast$ states are 
found to be originated
from the bare $N^\ast$ states [the second term of Eq.~(\ref{eq:pot})]
and there finds no meson-baryon molecular type of resonances, which
is generated dynamically from the pure meson-exchange processes.
\begin{center}
\tabcaption{ \label{tab1}  
The resonance pole positions $m_R$ [listed as $(\text{Re}~m_R, -\text{Im}~m_R)$]
 extracted from the EBAC-DCC model with the JLMS parameter set are compared 
with the values of 3- and 4-stars
nucleon resonances listed in the PDG~\citep{pdg}.
``---" for $P_{33}(1600)$, $P_{13}$ and $P_{31}$ indicates 
that no resonance pole has been found in the considered complex energy region, 
Re$(E)\leq 2000$ MeV and $-$Im$(E)\leq 250$ MeV.
}
\footnotesize
\begin{tabular*}{80mm}{ccccl}
\toprule &$m^0_{N^*}$&$m_R$& PDG  \\
         &(MeV)      &(MeV)& (MeV)  \\
\hline
$S_{11}$ &1800 &(1540,   191)&(1490 - 1530, \, 45 -   125)\\
         &1880 &(1642, \, 41)&(1640 - 1670, \, 75 - \, 90)\\
$P_{11}$ &1763 &(1357, \, 76)&(1350 - 1380, \, 80 -   110)\\
         &1763 &(1364,   105)&                            \\
         &1763 &(1820,   248)&(1670 - 1770, \, 40 -   190)\\
$P_{13}$ &1711 &---          &(1660 - 1690, \, 57 -   138)\\
$D_{13}$ &1899 &(1521, \, 58)&(1505 - 1515, \, 52 - \, 60)\\
$D_{15}$ &1898 &(1654, \, 77)&(1655 - 1665, \, 62 - \, 75)\\
$F_{15}$ &2187 &(1674, \, 53)&(1665 - 1680, \, 55 - \, 68)\\
$S_{31}$ &1850 &(1563, \, 95)&(1590 - 1610, \, 57 - \, 60)\\
$P_{31}$ &1900 &---          &(1830 - 1880,   100 -   250)\\
$P_{33}$ &1391 &(1211, \, 50)&(1209 - 1211, \, 49 - \, 51)\\
         &1600 & ---         &(1500 - 1700,   200 -  400)\\
$D_{33}$ &1976 &(1604,   106)&(1620 - 1680, \, 80 -   120)\\
$F_{35}$ &2162 &(1738,   110)&(1825 - 1835,   132 -   150)\\
         &2162 &(1928,   165)&                            \\       
$F_{37}$ &2138 &(1858,   100)&(1870 - 1890,   110 -   130)\\
\bottomrule
\end{tabular*}
\end{center}

We also find in Table~\ref{tab1} that
there exist two $P_{11}$ poles, $(1357,76)$ and $(1364,105)$, 
in the energy region close to the Roper resonance.
Here it is noted that recent analysis by the GWU and J\"{u}lich groups
have also reported the two $P_{11}$ poles in the same energy 
region\cite{said06,julich09}.
This is quite remarkable because, although the three approaches are 
based on completely different pictures for the Roper resonance,
all of the three show the two poles of the Roper resonance 
in the same complex energy region after fixing their model parameters by
fitting to the $\pi N$ scattering data. 
To conclude whether both of the two poles are the constituents 
of the Roper resonance, however, 
we need to make clear how these two poles affect the physics 
on the real energy axis.
This will be accomplished by investigating the magnitude
of the residue at the pole positions.

In Fig.~\ref{fig:3dplot}, we present the 3D plots of the $N^\ast$ pole
positions on the complex energy plane.
\end{multicols}

\ruleup
\begin{center}
\includegraphics[clip,width=13cm]{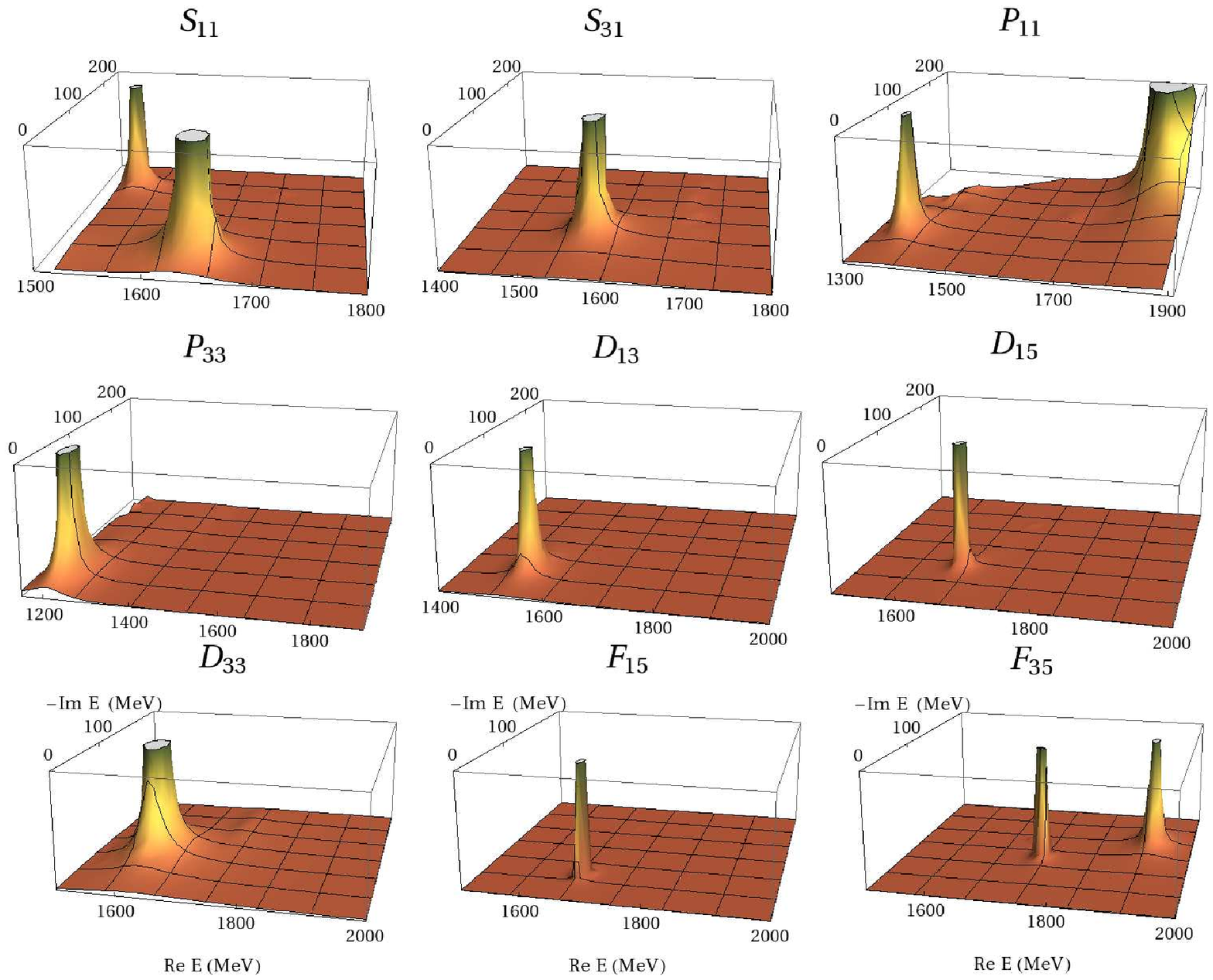}
\figcaption{\label{fig:3dplot}
The 3D plots of the $N^\ast$ pole positions on the complex $E$ plane.
The first two $P_{11}$ resonances are very close and cannot be shown explicitly.
}
\end{center}
\ruledown

%
%
\begin{multicols}{2}
\section{\label{sec:plan}Plans for future developments}
As a future work, the EBAC program has four main components
described in the following subsections.
\subsection{Extracting $N^\ast\to MB$ and $N^\ast\to \gamma^{(\ast)} N$ vertex functions}
In Sec.~{\ref{sec:pole}}, we have presented the $N^\ast$ pole positions extracted
from the current EBAC-DCC model.
As a next step we will move to the extraction of 
the $N^\ast\to MB$ and $N^\ast\to \gamma^{(\ast)} N$ vertex functions.
As mentioned in Sec.~{\ref{sec:pole}}, a proper, physically meaningful
$N^\ast$ vertex function is defined as a residue at the corresponding
$N^\ast$ pole, not at the Breit-Wigner mass.
Therefore we need to perform an analytic continuation of the scattering
amplitude as well as in the case of pole extraction.
\subsection{Analysis of photoproduction reactions and search for new $N^\ast$ states}
We plan to analyze photoproduction reactions of various final states
such as $KY$, $\eta N$, and $\pi\pi N$ in the same way as made
in the single pion photoproduction reactions.
As for these final states,
the amount of currently available data of the initial $\gamma N$ are
much larger than those of the initial $\pi N$.
Therefore we will use these photoproduction data to precisely determine
the parameters associated with $N^\ast\to KY$, $N^\ast\to \eta N$, 
and $N^\ast\to\pi\Delta,\rho N,\sigma N$
decay vertex functions.
Also, the analysis of these reactions is interesting 
because recent experiments have suggested new $N^\ast$ states which 
strongly couple to these reaction channels
but not to the dominant $\pi N$ channel\cite{mcnabb04,sumihama06}.
We could find a possible existence of such new $N^\ast$ states from
the analysis.
\subsection{Developing connection with hadron structure calculations}
All $N^\ast$ parameters in the EBAC-DCC model have hitherto been
determined by fitting to the data.
While this has greatly improved our knowledge of the spectrum,
it has not improved our understanding within QCD.
For this, it is necessary to develop a method
to connect our extracted $N^\ast$ information to the various hadron
structure calculations. 
This is crucial for reaching the final goal of EBAC.
Encouragingly, in that connection progress has been made in
constituent quark models\cite{inna08}
and Lattice QCD simulations\cite{alex05,lin08}.

\subsection{Upgrading EBAC-DCC model}
In parallel with the above investigations, we keep upgrading 
EBAC-DCC model.
At present the parameters associated with the hadronic interactions 
are determined by just analyzing the $\pi N\to \pi N$ scattering,
and the electromagnetic parameters are determined from the analysis
of the single pion photo- and electro-production reactions
in which the fixed value of the hadronic parameters are used.
However, we have seen that we need a combined, simultaneous analysis 
of the $\pi N$ and $\pi \pi N$ channels to obtain
more reliable model.
In the next step we will refine our model by performing
the combined analysis of the 
$\pi N\to\pi N,\pi\pi N$ and
$\gamma N\to\pi N,\pi\pi N$ reactions.

\begin{center}
\includegraphics[clip,width=5.5cm]{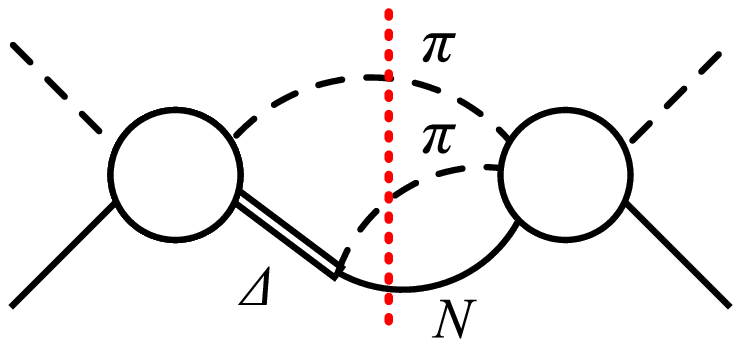}
\figcaption{
One example of diagrams including the three-body $\pi\pi N$ unitary cut.
The intermediate $\pi\pi N$ state at the red dotted line can be
on the mass-shell.
\label{fig:3-body-cut}
}
\end{center}

Also, we will incorporate all processes with the $\pi\pi N$
3-body unitary cut.
Some of the processes as shown in Fig.~\ref{fig:3-body-cut}
have not been included in our current model.
Such processes could be important for the meson production reactions
up to 2 GeV.
As far as we know, such complete treatment of the three-body
unitary cut has never been done before in the $N^\ast$ study,
and thus this improvement will make our model unique from
other dynamical models.\\

\acknowledgments{
The author would like to thank B.~Juli\'a-D\'{\i}az, T.-S.~H.~Lee, 
A.~Matsuyama, T.~Sato, and N.~Suzuki for their collaborations at EBAC.
This work used resources of the National Energy Research Scientific
Computing Center (NERSC) which is supported by the Office of Science of the 
U.S. Department of Energy under Contract No. DE-AC02-05CH11231.
}

\end{multicols}

\vspace{-2mm}
\centerline{\rule{80mm}{0.1pt}}
\vspace{2mm}

\begin{multicols}{2}

\end{multicols}

\end{document}